\documentclass[12pt]{article}
\usepackage[a4paper]{geometry}
\usepackage[myheadings]{fullpage}
\usepackage{fancyhdr}
\usepackage{graphicx, wrapfig, subcaption, setspace, booktabs}
\usepackage[T1]{fontenc}
\usepackage[font=small, labelfont=bf]{caption}
\usepackage{fourier}
\usepackage[protrusion=true, expansion=true]{microtype}
\usepackage[english]{babel}
\usepackage{sectsty}
\usepackage{amsmath,amsfonts,amssymb}
\usepackage{tocloft}
\graphicspath{ {figures/} }
\usepackage{array}
\usepackage{acronym} 
\usepackage{threeparttable}
\usepackage{adjustbox}
\usepackage{url}
\usepackage{siunitx}
\usepackage{makecell}
\usepackage[longnamesfirst,square]{natbib}

\usepackage{verbatim}

\begin{document}

\title{ \normalsize 
		\LARGE \textbf{Impact of AI-Triage on Radiologist Report Turnaround Time: Real-World Time-Savings and Insights from Model Predictions}
  }

\date{}


\author{
	  Yee Lam Elim Thompson (FDA/CDRH/OSEL) \\
        Jonathan Fergus (University of Chicago) \\
        Jonathan Chung (University of Chicago) \\
        Jana G. Delfino (FDA/CDRH/OSEL) \\  
        Weijie Chen (FDA/CDRH/OSEL) \\ 
        Gary M. Levine (FDA/CDRH/OPEQ) \\   
        Frank W. Samuelson (FDA/CDRH/OSEL)}

\maketitle

\newpage
\begin{acronym}[MPC] 
\acro{AI}{Artificial Intelligence}
\acro{FDA}{U.S. Food and Drug Administration}
\acro{CADt}{Computed Aided Triage and Notification}
\acro{PE}{Pulmonary Embolism}
\acro{CTPA}{CT pulmonary angiography}
\acro{TAT}{Turnaround Time}
\acro{PACS}{Picture Archiving and Communication System}
\acro{CI}{Confidence Interval}
\acro{SD}{Standard Deviation}
\end{acronym}

\sectionfont{\scshape}



\newpage
\section{Abstract}
\noindent\textbf{Objective}: To quantify the impact of workflow parameters on time-savings in report turnaround time due to an AI-triage device that prioritized pulmonary embolism (PE) in chest CT pulmonary angiography (CTPA) exams.

\noindent\textbf{Methods}: This retrospective study analyzed 11,252 adult CTPA exams conducted for suspected PE at a single tertiary academic medical center. Data was divided into two periods: pre-AI and post-AI. 
For PE-positive exams, turnaround time (TAT)—defined as the duration from patient scan completion to the first preliminary report completion—was compared between the two periods. Time-savings were reported separately for work-hour and off-hour cohorts. To characterize radiologist workflow, 527,234 records were retrieved from the Picture Archiving and Communication System (PACS) and workflow parameters such as exam inter-arrival time and radiologist read-time extracted. These parameters were input into a computational model to predict time-savings following deployment of an AI-triage device and to study the impact of workflow parameters. 

\noindent\textbf{Results}: The pre-AI dataset included 4,694 chest CTPA exams with 13.3\% being PE-positive. The post-AI dataset comprised 6,558 exams with 16.2\% being PE-positive. 
The mean TAT for pre-AI and post-AI during work hours are 68.9 [95\% CI" 55.0, 82.8] and 46.7 [38.1, 55.2] minutes respectively, and those during off-hours are 44.8 [33.7, 55.9] and 42.0 [33.6, 50.3] minutes. Clinically-observed time-savings during work hours (22.2 [95\% CI: 5.85, 38.6] minutes) were significant ($p$=0.004), while off-hour  (2.82 [-11.1, 16.7] minutes) were not ($p$=0.345). Observed time-savings aligned with model predictions (29.6 [95\% range: 23.2, 38.1] minutes for work hours; 2.10 [1.76, 2.58] minutes for off-hours). 

\noindent\textbf{Discussion}: Consideration and quantification of the clinical workflow contributes to the accurate assessment of the expected time-savings in report turnaround time following deployment of an AI-triage device.

\newpage
\section{Keywords}
\begin{itemize}
    \item AI-enabled triage medical devices
    \item Radiologist workflow
    \item Time-savings in report turnaround time
\end{itemize}

\newpage
\section{Summary Sentence}
Clinically-observed time-savings in report turnaround time from an AI-triage varied based on radiologist workflow and aligned with predictions from a computational model, underscoring the importance of clinical workflow in determining effectiveness of AI-triage devices.

\newpage
\section{Introduction}
\label{introduction}



A Computer-Aided Triage and Notification (CADt) device, commonly known as an AI-triage device, is software intended to identify patient exams containing radiological medical images with positive, time-sensitive findings and prioritize the study in a radiologist's reading queue. This allows radiologists to review imaging studies and communicate urgent findings, potentially leading to earlier treatment and improved patient outcomes. Pulmonary embolism (PE) is one such critical condition, where timely diagnosis is key to better patient care~\citep{Wood2002-ho}. Multiple studies have examined the performance of AI-triage devices for PE-positive CTPA exams~\citep{Batra2022-og,Katzman2023-bx,Weikert2020-es,Zaazoue2023-bq,Langius-Wiffen2023-tz,Ebrahimian2022-ls}. At the time of writing, eight AI-triage devices have been FDA-cleared for prioritizing CTPA exams for PE, incidental PE (iPE), and central PE, all reporting sensitivities and specificities over 80\% in their FDA 510(k) summaries~\citep{FDAdatabase1, FDAdatabase2}.

However, studies have reported conflicting results regarding the time-savings of AI-triage in radiologist report turnaround time (TAT) for PE-positive exams. Schmuelling \text{et al.} found a non-significant mean time-savings of 7 minutes in report TAT after AI deployment ($p$ = 0.911)~\citep{Schmuelling2021}. In contrast, Batra \textit{et al.} reported a significant 12-minute report-TAT time-savings for PE-positive exams~\citep{Batra2023}, and Rothenberg \text{et al.} saw mean report TAT drop from 52.1 minutes to 44.0 minutes ($p$ = 0.11)\citep{Rothenberg2023-wq}. For iPE, Topff \textit{et al.} reported a drastic reduction in median report TAT from 7,772 minutes to 148 minutes post-AI-triage~\citep{Topff2023}, while Savage \text{et al.} found no significant change in report TAT ($p$ = 0.26)~\citep{Savage2024}.

We previously developed QuCAD, a computational model that simulates and predicts time-savings in report TAT following AI-triage device deployment~\citep{Thompson2022-vk}. The simulation software was analytically verified by queueing theory~\citep{Thompson2024-ce}. QuCAD incorporates information about an AI-triage device, such as true and false-positive fractions, along with clinical workflow parameters (disease prevalence, exam inter-arrival time, radiologist read-time, and the number of available radiologists), and estimates the average time-savings for the diseased population after the deployment of the AI-triage device.

The aims of this study are to 1) quantitatively characterize the radiological workflow, 2) use QuCAD to predict time-savings in report TAT for PE-positive exams after deployment of an AI triage device, and 3) investigate the effects of workflow parameters on the time-saving observed following deployment of an AI-triage device. All aims concern the duration between patient scan completion time and the first preliminary report completion time for PE-positive CTPA exams. 

\section{Materials and Methods}
\label{method}

\subsection*{Study design and participants}
This retrospective single-center study aggregated and summarized metadata with no Personal Identifiable Information. The study was reviewed by the University of Chicago  Institutional Review Board (IRB) as IRB22-0704 and determined exempt due to minimal risk. 

\begin{figure*}[h!]
  \includegraphics[width=0.7\linewidth]{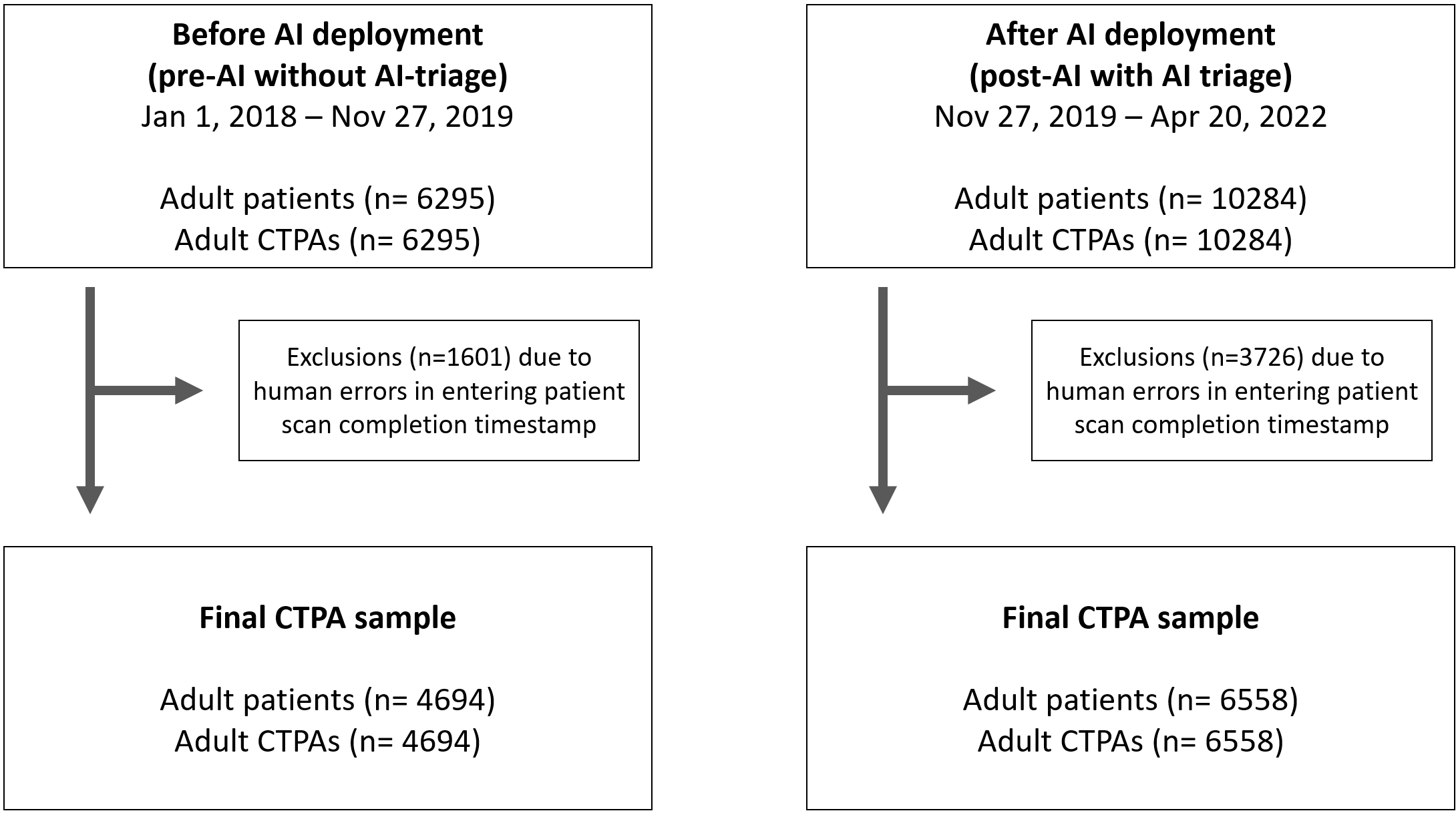}
  \centering
  \caption{Flowchart depicting the number of adult chest CTPA exams before and after the AI-triage device was implemented. AI = Artificial Intelligence; CTPA = CT Pulmonary Angiography.}
  \label{fig:flowchart}
\end{figure*}

Between January 1, 2018 and April 20, 2022, 16,579 consecutive adult chest CTPA exams were identified by Nuance mPower Clinical Analytics (mPower), a radiology reporting software. For each exam, patient scan completion time, the anonymized ID of the first reviewing radiologist, radiologist sign-off time, and the final diagnosis by a senior staff radiologist were extracted from the radiology report. Report TAT is the time duration between patient scan completion and the first preliminary report completion. 5,327 exams with negative TATs were excluded from the analysis (Figure~\ref{fig:flowchart}). Negative TATs result when the scan completion time (manually entered) is after the report completion time (automatically captured). This leaves a final sample of 11,252 exams, each corresponding to a unique patient. TATs for PE-positive exams were analyzed separately for work hours (weekdays, 8 am–5 pm) and off-hours (weekends, nights, holidays).

Radiologists who reviewed CTPA exams also read non-CTPA exams, which were not included in the mPower dataset. To extract additional workflow parameters, we used timestamp data of CTPA and non-CTPA exams from the Picture Archiving and Communication System (PACS). Between January 1, 2018, and April 20, 2022, PACS recorded 958,089 entries, each documenting the radiologist who closed an exam and the corresponding case-closure time. We limited the workflow parameter estimation to the 527,234 entries that included radiologists who also read CTPA exams.

\subsection*{AI-enabled triage algorithm}
During the post-AI period, CTPA exams were processed by an FDA-cleared AI-triage device (BriefCase by Aidoc Medical)\footnote{Aidoc Medical was not involved in this study and did not provide any financial support for this study.}. The device is intended to identify suspected PE in CTPA exams, with a reported sensitivity of 90.6\% [95\% confidence interval (CI): 82.2\%, 95.9\%] and specificity of 89.9\% [95\% CI: 82.2\%, 95.1\%]~\citep{510ksummary}.

\subsection*{Reference Standard}
Each radiologist report included a pick list with the options ``PE: Positive, Negative, Indeterminate''. Only ``PE:Positive'' was considered as PE-positive, and both ``PE:Negative'' or ``PE:Indeterminate'' were considered as non-PE-positive. Each report included an initial diagnosis and a final diagnosis, with the final diagnosis serving as the reference standard. 

\subsection*{Turnaround time and time-savings}
Radiologist report TAT is the time from scan completion to when the first radiologist (resident or staff radiologist) signs off. Time-savings is the difference in report TAT before and after AI-triage deployment. For each cohort (work-hour and off-hour), we reported the mean report TAT for PE-positive exams before and after AI-triage deployment, as well as their difference (i.e. time-savings) for PE-positive exams.

\subsection*{Workflow parameters for QuCAD model}\label{sec:workflowParams}
The radiologist workflow was quantitatively modeled by several parameters, including disease prevalence, exam inter-arrival time, the number of radiologists available to review images, their read-time, and the TPF and FPF of the AI-triage device.

QuCAD assumes the AI-triage device reviews all images in the reading queue. However, in practice, radiologists who review CTPA exams also handle studies that the AI does not analyze in the clinical setting because these exams are out of scope for the AI (e.g. chest X-rays), which we refer to as ``non-chest/CT'' studies. Therefore, the reading queue consists of three exam types: PE-positive CTPA exams, non-PE-positive CTPA exams, and non-chest/CT exams.

Two prevalence rates are reported. The first is the prevalence commonly defined as the ratio of PE-positive exams to the total number of CTPA studies. However, to account for non-chest/CT studies in the reading queue, we also report the ratio of PE-positive exams with respect to all studies in the queue, as this is used as the disease prevalence input for QuCAD.

Since QuCAD assumes exams arrive following a Poisson distribution~\citep{Stewart2009-cz}, the mean inter-arrival time between two consecutive exams follows an exponential distribution~\citep{Budgett2018-cl}, where the mean equals the inverse of the Poisson arrival rate. We checked the exponential assumption by fitting an exponential curve to the daily inter-arrival time distribution. To characterize the overall mean inter-arrival time across all days, we presented a distribution of the mean exam inter-arrival times and fitted a normal distribution to describe it via the mean and 1$\sigma$ (68\%) range of the distribution.  These results were reported separately for work-hour and off-hour cohorts. QuCAD was run iteratively for a range of mean inter-arrival times covering the 1$\sigma$ ranges of both cohorts.

Timestamp data cannot be used to determine the number of radiologists (residents and attendings) readily available to review CTPA exams at any given time point. However, according to the radiologists involved in this study, the department typically has three to four radiologists on-site to review CTPA exams. Therefore, we ran QuCAD iteratively with two to five radiologists.

PACS captured exam closing, but not opening time, making it impossible to directly extract case reading time. We indirectly estimated reading time by grouping case-closure times for individual radiologists and calculating the time between consecutive case closures; this inter-case-closure time served as a surrogate for the reading time. This approximation assumes radiologists are reading cases consecutively which is an accurate assumption for residents at our institution, but not senior staff radiologists (who also often handle non-image-related tasks). Therefore, read-time results are limited to residents.  Additionally we excluded cases with inter-case-closure times exceeding one hour and omitted cases from days when the resident handled fewer than 30 cases.  

QuCAD accommodates different reading rates for diseased and non-diseased populations. To model this, we grouped read-times by exam type (PE-positive, non-PE-positive, and non-chest/CT) for each resident and calculated the average read-time for each type. Similar to the arrival pattern, QuCAD also assumes an exponential distribution for read-time. We checked this assumption by fitting an exponential curve to a read-time distribution of each resident who reviewed at least 10 exams of a given exam type.

The mean read-time from PE-positive exams was used as the read-time for the diseased population in QuCAD. For non-diseased cases, an effective mean read-time was calculated by weighting the mean read-times of non-PE-positive and non-chest/CT exams based on their proportions in the PACS data. QuCAD assumes that radiologist read-time performance is not affected by the use of AI. Since the impact of AI triage on read-times was unknown, only pre-AI data was used to estimate read-times for QuCAD inputs.

We used the reported sensitivity from the FDA 510(k) summary~\citep{510ksummary} as the model input for AI's TPF. However, the FPF for model input was adjusted by the ratio between non-chest/CT and non-PE-positive exams because the reported specificity was based on a dataset with only CTPA exams, yet the reading queue includes non-chest/CT exams. 
\begin{equation*}\label{eqn:adjustedFPF}
\begin{split}
    \text{FPF}_{\text{adjusted}} &= \frac{N_{\text{FP}}}{N_{\text{non-PE-positive}} + N_{\text{non-chest/CT}}} \\
                                 &= \frac{\text{FPF}_{\text{reported}}}{1 + N_{\text{non-chest/CT}}/N_{\text{non-PE-positive}}}\\
                                 &= \frac{1-\text{Sp}_{\text{reported}}}{1 + N_{\text{non-chest/CT}}/N_{\text{non-PE-positive}}}
\end{split}
\end{equation*}
where $N_{\text{FP}}$, $N_{\text{non-PE-positive}}$, and $N_{\text{non-chest/CT}}$ are the number of false-positive, non-PE-positive, and non-chest/CT exams respectively. $\text{Sp}_{\text{reported}}$ is the specificity from the 510(k) summary, and $\text{FPF}_{\text{adjusted}}$ is the adjusted FPF fed to QuCAD. This adjustment simulates a reading queue as if the AI-triage analyzes non-chest/CT exams and accurately classifies them as true-negatives. It mirrors the clinical use of the AI-triage device, where AI-positive cases were prioritized over all other exams.

\subsection*{Workflow simulation}
We simulated two workflow settings. First, QuCAD was run using the disease prevalence and mean read-times extracted from the data and varying the number of radiologists and mean exam inter-arrival times. For each simulation setting, 100 trials each with 100,000 simulated patients were conducted to calculate the mean time-savings for PE-positive exams. We then compared the predicted and clinically-observed time-savings and investigated   impact due to changes in the workflow parameters. Second, we fit a receiver operating characteristic (ROC) curve using a bi-normal assumption based on the sensitivity and specificity in the 510(k) summary. To explore how adjusting TPF and FPF might impact time-savings, we ran QuCAD across a range of FPF and TPF along the ROC curve, assuming three radiologists and a mean exam inter-arrival time from the work-hour cohort.

\subsection*{Statistical analysis}
Descriptive statistics for CTPA exams are presented as frequency with percentage for categorical variables and as mean $\pm$ standard deviation (SD) for continuous variables. The 29-month post-AI period had more exams than the 23-month pre-AI period, and the standardized differences are reported. The clinically-observed mean TAT and time-savings among PE-positive exams are reported with their 95\% CIs.

When fitting exponential curves to the distributions of inter-arrival time and per-resident reading time, the coefficient of determination ($R^2$) was used to describe the goodness-of-fit. Both mean and SD of $R^2$ are reported. The QuCAD-predicted mean time-savings for PE-positive exams is reported with a 2-tailed 95\% range based on 100 simulation trials. The 95\% range from QuCAD are compared to the 95\% CIs of the clinically-observed time-savings to determine if QuCAD's time-savings agrees with the clinically-observed time-savings.

\section{Results}
\subsection*{Participant Characteristics}
The final dataset included 4,694 CTPA exams before AI-triage deployment (mean age, 53.7 years $\pm$ 18.4; 2990 female, 1704 male) and 6,558 exams after deployment (mean age, 54.9 years $\pm$ 18.2; 3981 female, 2572 male, 5 unknown). Small standardized differences were observed in the characteristics of exams between the pre- and post-AI periods (Table~\ref{table:mPowerCharacteristics}).

\begin{table}[!ht]\footnotesize
	\centering
    \begin{threeparttable}
	\caption{Participant and CTPA examination characteristics. Data are numbers of participants with percentages in parentheses unless otherwise indicate. The statistical analyses are based on the sample proportion, not counts. Standardized differences are reported for both continuous and categorical variables. AI = Artificial Intelligence; PE = Pulmonary Embolism; CTPA = CT Pulmonary Angiography; SD = Standard Deviation.}
	\label{table:mPowerCharacteristics}
	\begin{tabular}{lr@{ }lr@{ }lr@{}}
	\toprule
	Characteristics & Pre-AI & (n=4694) & Post-AI & (n=6558) & Standardized Difference\\
	\midrule
	Sex & & \\
    \hspace{1cm} Female  & 2990 & (63.7) & 3981 & (60.7) & -0.044 \\
    \hspace{1cm} Male    & 1704 & (36.3) & 2572 & (39.2) & 0.043 \\
    \hspace{1cm} Unknown & 0    & (0.0)  &    5 & (0.08) & 0.028 \\
    \midrule
	Age (year)\tnote{*} & 53.7 &$\pm$ 18.4 & 54.9 & $\pm$ 18.2 & 0.045 \\
    \midrule
    Examination Location & & \\
    \hspace{1cm} Emergency department & 1368 & (29.1) & 1796 & (27.4) & -0.028 \\
    \hspace{1cm} Inpatient            & 2806 & (59.8) & 4294 & (65.5) & 0.083 \\
    \hspace{1cm} Outpatient           &  520 & (11.1) & 468  & (7.14) & -0.097 \\
    \midrule
    PE diagnosis\tnote{$\ddagger$} & & \\
    \hspace{1cm} Positive      &  623 & (13.3) & 1060 & (16.2) & 0.058 \\
    \hspace{1cm} Negative      & 3894 & (83.0) & 5194 & (79.2) & -0.068 \\
    \hspace{1cm} Indeterminate &  177 &  (3.8) &  304 &  (4.6) & 0.030 \\
    \midrule
    PE location\tnote{\textdagger} & & \\
    \hspace{1cm} Main          & 114 & (18.3) & 203 & (19.2) & 0.015 \\
    \hspace{1cm} Lobar         & 148 & (23.8) & 244 & (23.0) & -0.012 \\
    \hspace{1cm} Segmental     & 220 & (35.3) & 388 & (36.6) & 0.019 \\
    \hspace{1cm} Subsegmental  & 127 & (20.4) & 203 & (19.2) & -0.022 \\
    \hspace{1cm} Indeterminate &  14 &  (2.3) &  22 &  (2.1) & -0.008 \\
    \midrule
    PE chronicity\tnote{\textdagger} & & \\
    \hspace{1cm} Acute         & 465 & (74.6) & 839 & (79.2) & 0.076 \\
    \hspace{1cm} Chronic       &  99 & (15.9) & 152 & (14.3) & -0.031 \\
    \hspace{1cm} Indeterminate &  59 &  (9.5) &  69 &  (6.5) & -0.077 \\
    \midrule
    PE multiplicity\tnote{\textdagger} & & \\
    \hspace{1cm} Single        & 190 & (30.5) & 336 & (31.7) & 0.018 \\
    \hspace{1cm} Multiple      & 412 & (66.1) & 699 & (65.9) & -0.003 \\
    \hspace{1cm} Indeterminate &  21 &  (3.4) &  25 &  (2.4) & -0.043 \\
    \midrule    
    Number of reader & & \\
    \hspace{1cm} All titles          & 63 &  (100) & 72 &  (100) & -     \\
    \hspace{1cm} Resident            & 41 & (65.1) & 41 & (56.9) & -0.118 \\
    \hspace{1cm} Staff               & 15 & (23.8) & 20 & (27.8) & 0.064 \\
    \hspace{1cm} L1 Fellow           &  2 &  (3.2) &  1 &  (1.4) & -0.085 \\
    \hspace{1cm} Emergency Physician &  5 &  (7.9) & 10 & (13.9) & 0.136 \\
	\bottomrule
	\end{tabular}
    \begin{tablenotes}
    \item[*] Data are means $\pm$ SDs.
    \item[$\ddagger$] PE diagnosis refers to the final diagnoses reviewed by senior staff radiologists.
    \item[\textdagger] Only PE-positive exams were included in this category. Percentages are with respect to the number of PE-positive exams.
    \end{tablenotes}
    \end{threeparttable}
\end{table}

\subsection*{Turnaround time and time-savings}
Before the deployment of AI-triage, the clinically-observed mean TATs for PE-positive exams are 68.9 [95\% CI: 55.0, 82.8] and 44.8 [33.7, 55.9] minutes during work hours and off-hour respectively. After the deployment, the mean TATs are 46.7 [38.1, 55.2] and 42.0 [33.6, 50.3] minutes (Table~\ref{tab:tatdata}). Time-savings of 22.2 minutes [5.85, 38.6] were observed during work hours ($p$ = 0.004), while no significant savings were found during off-hours (2.82 [-11.1, 16.7], $p$ = 0.345). 

\begin{table}[!ht]\footnotesize
	\centering
    \begin{threeparttable}
    \begin{tabular}{c|p{25mm}p{25mm}|p{25mm}c|p{28mm}}
	\toprule

               & Pre-AI TAT [min] \newline [95\% CI] \newline (2.5, 50, 97.5-th) &Post-AI TAT [min] \newline [95\% CI] \newline (2.5, 50, 97.5-th) & Observed time-\newline savings (95\% CI) & $p$-value & Predicted time-\newline savings (95\% range) \\               
	\midrule
    Work-hour  & 68.9 \newline [55.0, 82.8] \newline (1.93, 35.4, 425) & 46.7 \newline [38.1, 55.2] \newline (2.46, 22.4, 199)  & 22.2 \newline [5.85, 38.6] &  0.004  & 29.6 \newline [23.2, 38.1] \\     
    \specialrule{0.1pt}{0pt}{0pt}
    Off-hour   & 44.8 \newline [33.7, 55.9] \newline (1.92, 24.7, 175) & 42.0 \newline [33.6, 50.3] \newline (1.30, 26.0, 186) & 2.82 \newline [-11.1, 16.7] &  0.345 &  2.10 \newline [1.76, 2.58] \\    
	\bottomrule
    \end{tabular}
    \end{threeparttable}
	\caption{Rows correspond to the results during work hours and off-hour respective. First two columns are the mean report TAT for PE-positive exams before and after AI-triage deployment with their 95\% CIs and the 2.5-th, 50-th, and 97.5-th percentiles. Third column is the clinically-observed time-savings with their 95\% CIs, followed by the $p$-value (forth column), indicating how likely the time-savings from clinical data are to be greater than zero. Fifth column is the predicted time-savings from QuCAD with their 95\% range. CI = Confidence Interval; AI = Artificial Intelligence; TAT = Turnaround Time; PE = Pulmonary Embolism.}
	\label{tab:tatdata}
\end{table}

\subsection*{Workflow parameters for QuCAD model}

The prevalence of PE-positive exams among all CTPA exams was 15.0\% (1,683/11,252). Prevalence of PE-positive cases among all exams reviewed by radiologists who also reviewed CTPA exams was 1,683 out of 527,234 (0.319\%), which was used as the input prevalence for QuCAD.

\begin{figure}
\centering
\begin{subfigure}[b]{1\textwidth}
   \includegraphics[width=0.75\linewidth]{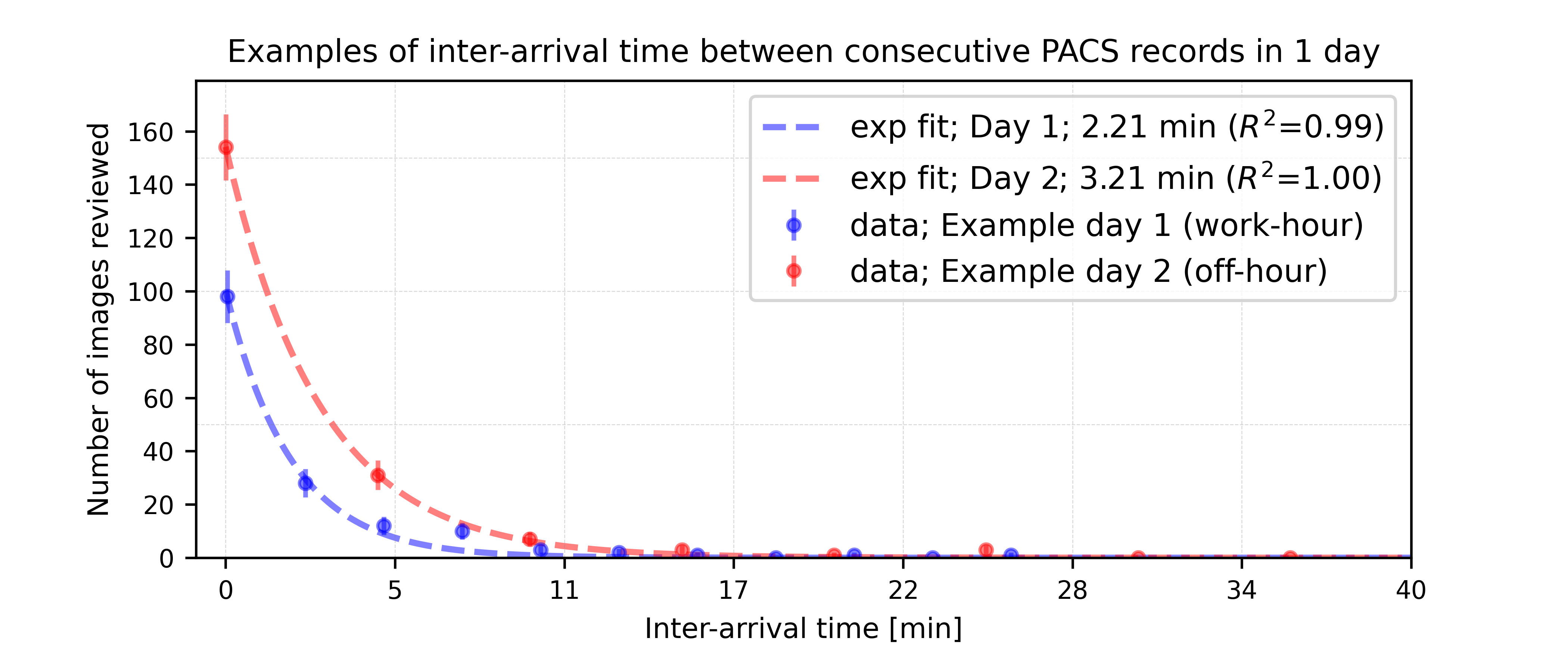}
   \centering
   \caption{Exam inter-arrival time distributions for two example days. Example day 1 includes data during work hours (blue dots with Poisson uncertainty), whereas day 2 includes data during off hours (red dots). Data from each day was described by an exponential curve (dashed lines). The mean inter-arrival times obtained from the fits are stated in legend. }
   \label{fig:interArrivalExamples} 
\end{subfigure}

\vspace{0.5mm}

\begin{subfigure}[b]{1\textwidth}
   \includegraphics[width=0.75\linewidth]{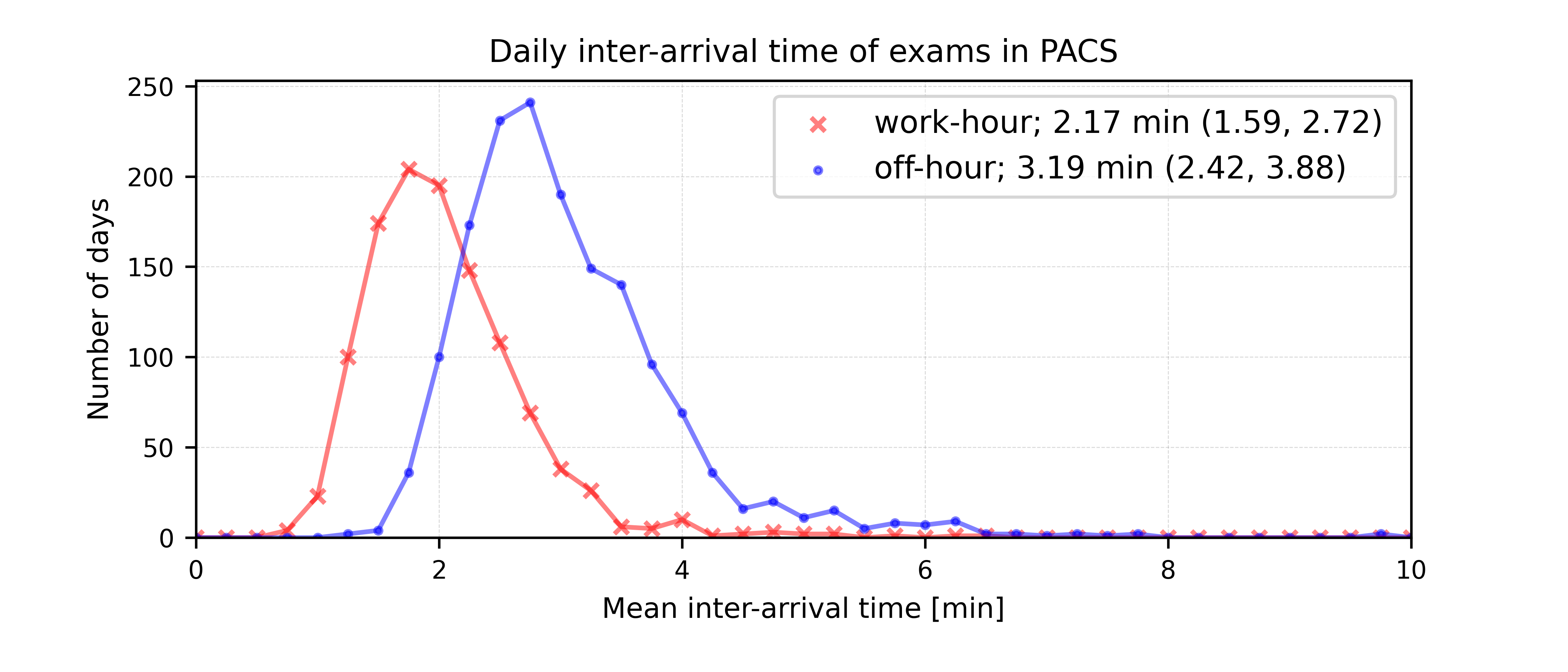}
   \centering
    \caption{Distributions of mean inter-arrival time from work-hour (red crosses) and off-hour (blue dots) cohorts. The average mean inter-arrival time and its 68\% range are shown in legend.}
   \label{fig:interArrivalHist}
\end{subfigure}

\vspace{0.5mm}

\begin{subfigure}[b]{1\textwidth}
    \centering
   \includegraphics[width=0.9\linewidth]{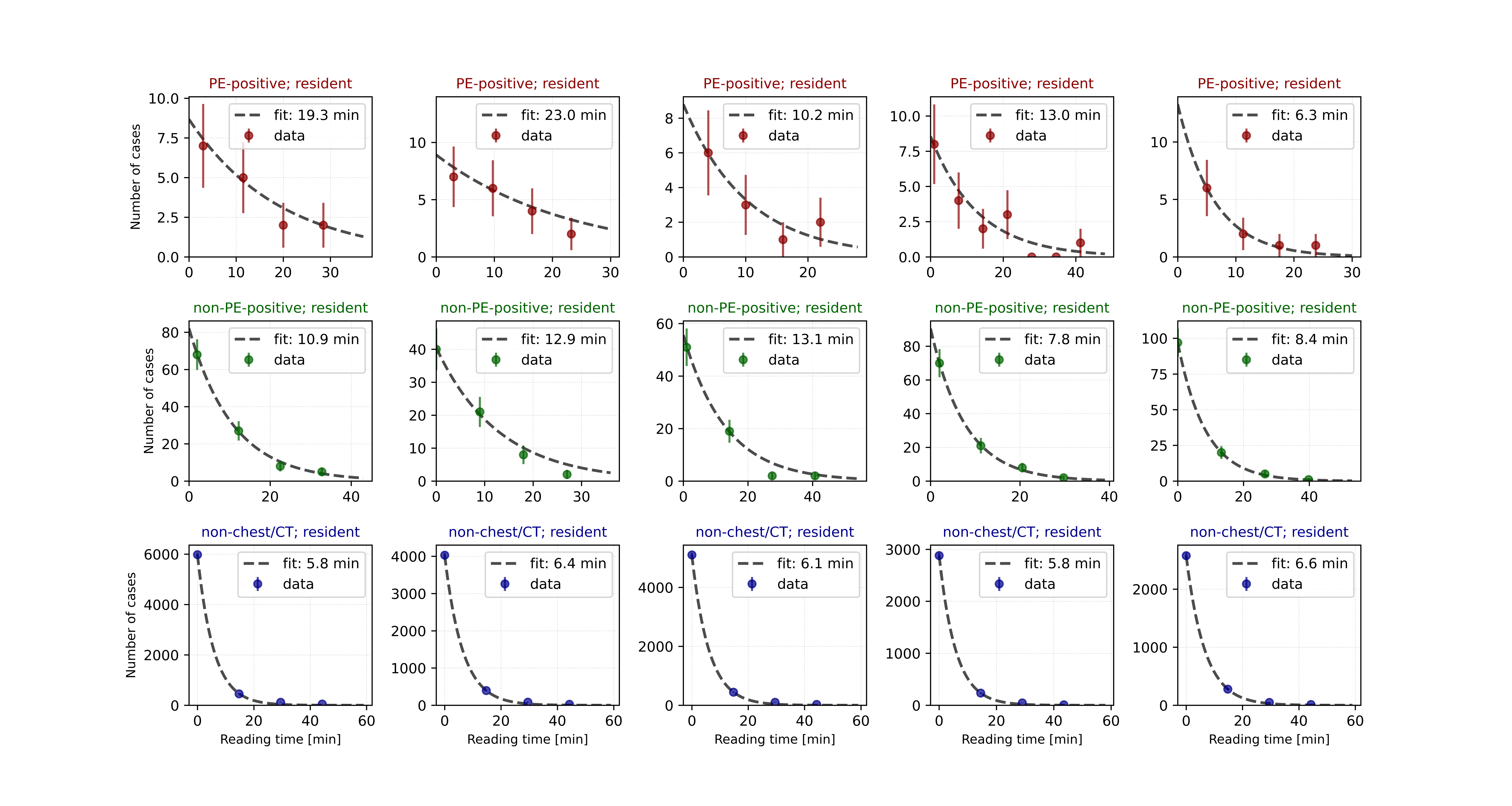}
   \centering
   \caption{Example read-time distributions per resident for PE-positive exams (top in red), non-PE-positive exams (middle in green), and non-chest/CT exams (bottom in blue). Each subplot includes data points from a resident and an exponential fit (dashed line) with the mean read-time in the legend. PE = Pulmonary Embolism.}
   \label{fig:expRateRead}
\end{subfigure}

\caption{Results of workflow parameters from data for modeling.}
\end{figure}

Figure~\ref{fig:interArrivalExamples} shows the distributions of exam inter-arrival times and their exponential fits for two randomly selected days. Each day includes two mean inter-arrival times (work-hour and off-hour cohorts), along with their corresponding $R^2$ values. This process was repeated for every day yielding a mean $R^2$ of 0.995 $\pm$ 0.005 for work hours and 0.995 $\pm$ 0.010 for off-hours.

The distributions of mean inter-arrival time for work-hour and off-hour cohorts are shown in Figure~\ref{fig:interArrivalHist}. Comparing the work-hour distribution with a mean of 2.17 minutes [68\% range: 1.59, 2.72] to that of off-hours (3.19 minutes [68\% range: 2.42, 3.88]) shows that a higher patient flow is observed during work hours. When generating time-saving predictions, QuCAD was run for mean inter-arrival times between 1.25 and 4.0 minutes, covering the variability from both distributions. 

Figure~\ref{fig:expRateRead} presents example distributions of read-time from five residents for PE-positive, non-PE-positive, and non-chest/CT exams. By performing an exponential fit to each read-time distribution, the average $R^2$ are 0.724 $\pm$ 0.246, 0.922 $\pm$ 0.142, 0.997 $\pm$0.016 for PE-positive, non-PE-positive, and non-chest/CT exams respectively. The mean read-time, along with the minimum and maximum, are reported in Table~\ref{tab:readTime}.

\begin{table}[!ht]\footnotesize
	\centering
    \begin{threeparttable}
	\begin{tabular}{c|ccc}
    \toprule
    Exam type       & Number of residents & Average mean read-time [min] & min, max range [min] \\ 
    \midrule
	PE-positive     &                  15 &                         12.1 &            8.2, 16.1 \\ 
    non-PE-positive &                  31 &                         11.4 &            7.7, 15.4 \\ 
    non-chest/CT    &                  34 &                          6.1 &            4.9,  8.9 \\ 
    \bottomrule
    \end{tabular}
    \end{threeparttable}
	\caption{Descriptive statistics for the mean read-times of residents. The first column indicates the number of residents who reviewed at least 10 exams of the corresponding exam type. The second column is the average mean read-time, and the third column is the minimum and maximum range. PE = Pulmonary Embolism.}
	\label{tab:readTime}
\end{table}

A mean read-time of 12.1 minutes was used as the QuCAD input for the diseased population. For the non-diseased population, an effective mean read-time of 6.15 minutes, calculated based on the relative proportions between the non-PE-positive and non-chest/CT exams, was used as model input.

In the first workflow setting, the reported sensitivity of 90.6\%~\citep{510ksummary} was used as the TPF input of the AI triage device. With the reported specificity of 89.9\%~\citep{510ksummary}, an adjusted FPF of 0.206\% was calculated and used as the model input. 

\begin{figure*}[h!]
  \includegraphics[width=0.6\linewidth]{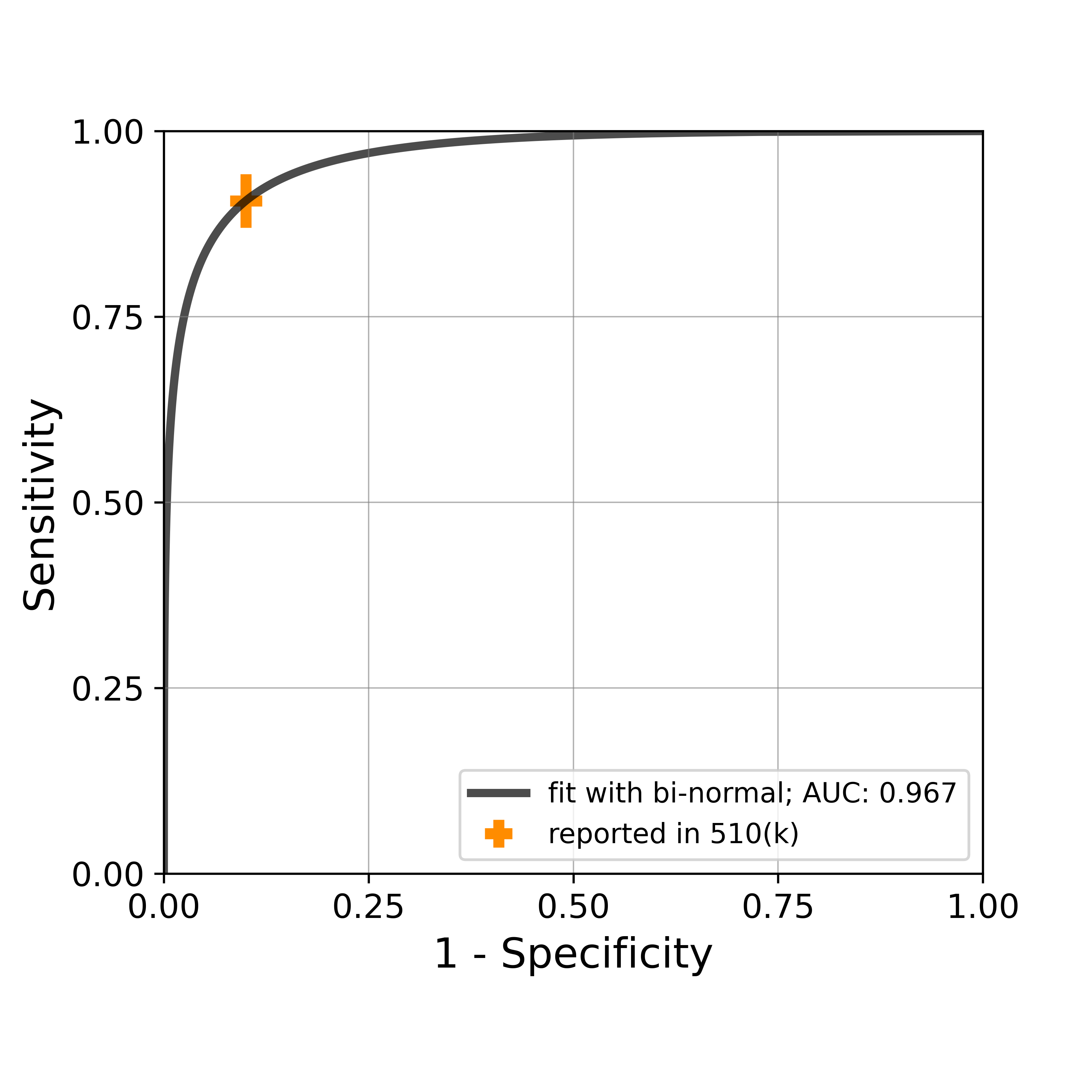}
  \centering
  \caption{An ROC curve using the sensitivity and specificity reported in the FDA 510(k) summary under a bi-normal assumption. ROC = Receiver Operating Characteristic; AUC = Area Under the ROC Curve; FDA = U.S. Food and Drug Administration.}
  \label{fig:AIROC}
\end{figure*}

For the second workflow setting, an ROC curve was fit under a bi-normal assumption (Figure~\ref{fig:AIROC}) based on the reported sensitivity and specificity. 1000 (FPF, TPF) points along the ROC curve were used as model input; these FPFs were also adjusted for the presence of non-chest/CT exams.

\subsection*{Workflow simulation}
\begin{figure}[h!]
\centering

\begin{subfigure}[b]{1\textwidth}
  \includegraphics[width=0.85\linewidth]{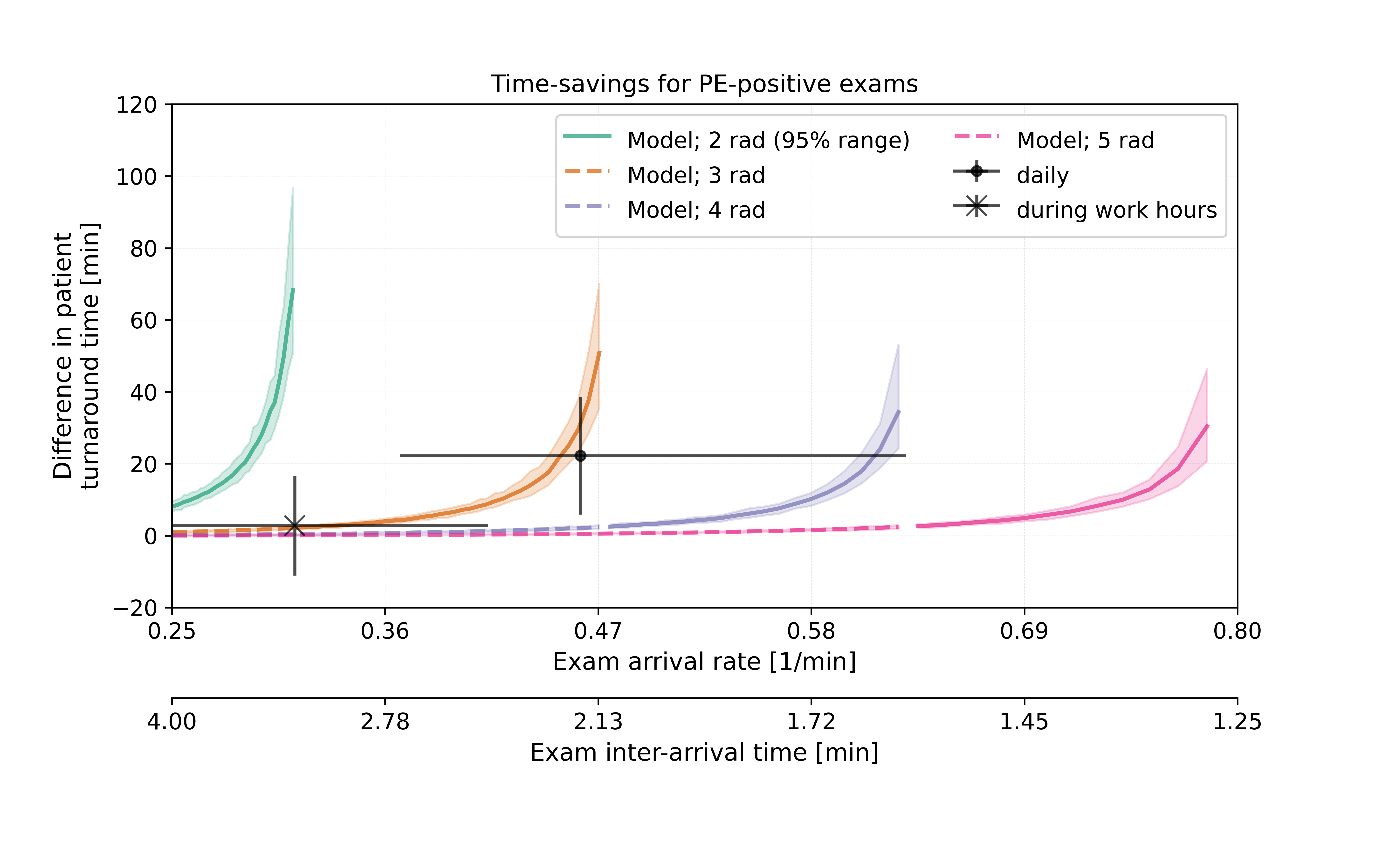}
  \centering
  \caption{Comparison of time-savings for PE-positive exams between QuCAD predictions and clinical data. The colored curves represent model predictions with two to five radiologists. The solid portion of the curves represent the maximum time-savings possible as a function of mean inter-arrival time, whereas the dashed potions indicate scenarios with too many radiologists available for the number of incoming exams. The black data points represent time savings from clinical data; their uncertainties in mean inter-arrival times are the 1$\sigma$ (68\%) ranges in Figure~\ref{fig:interArrivalHist}. PE = Pulmonary Embolism.}
  \label{fig:modelVSdata}
\end{subfigure}

\vspace{0.5mm}

\begin{subfigure}[b]{1\textwidth}
  \includegraphics[width=0.76\linewidth]{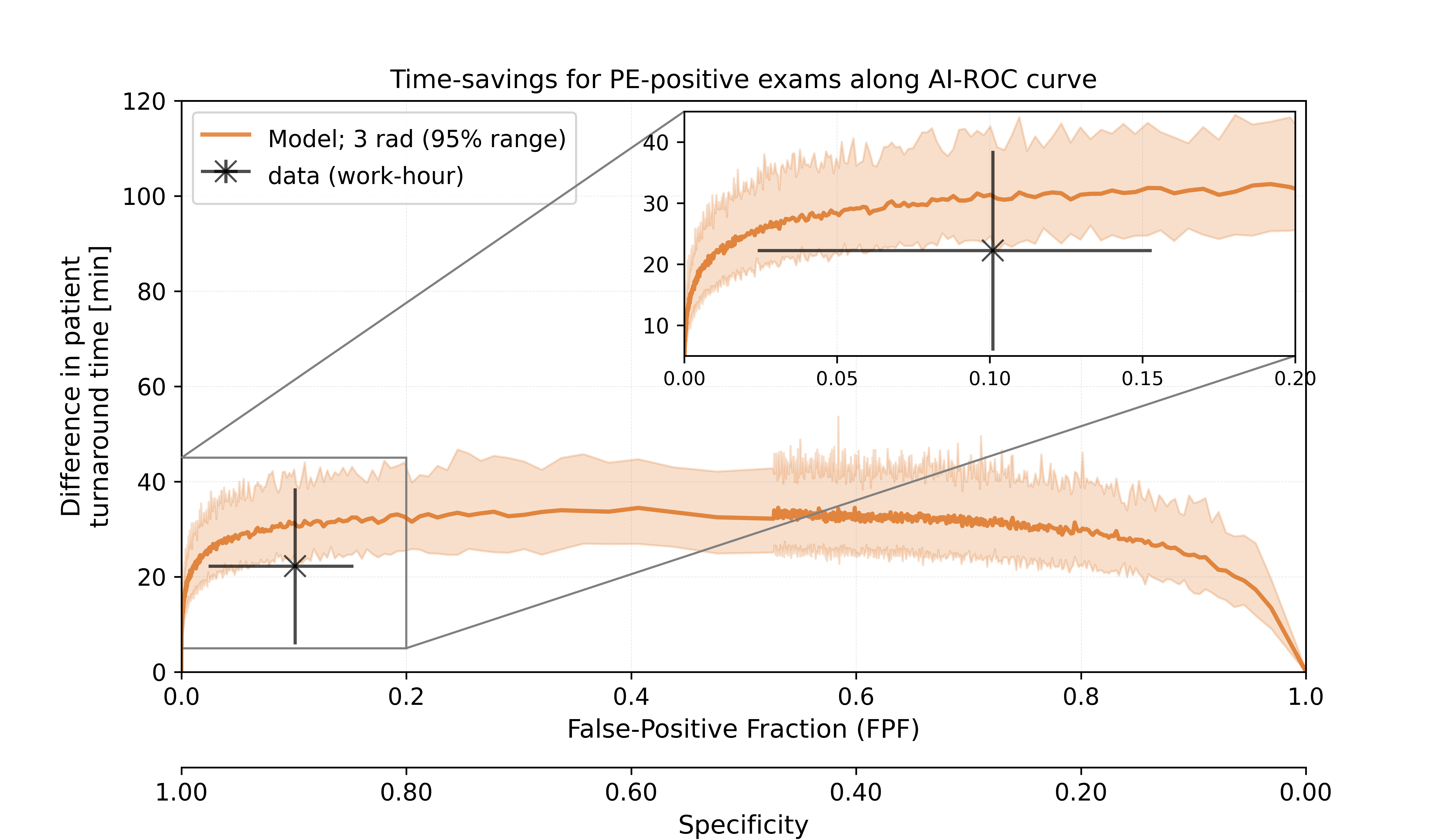}
  \centering
  \caption{Time-savings for PE-positive exam in the work-hour cohort along the AI-ROC curve. The curve represents the predicted time-savings with the 95\% range (shaded area), whereas the data point is the clinically-observed time-savings in the work-hour cohort. The FPF uncertainty of the data point was obtained from the FDA 510(k) summary~\citep{510ksummary}. PE = Pulmonary Embolism; AI = Artificial Intelligence; ROC = Receiver Operating Characteristic; FPF = False-Positive Fraction; FDA = U.S. Food and Drug Administration.}
  \label{fig:model_TPF}
\end{subfigure}

\caption{Results of clinically-observed and model-predicted time-savings.}
\end{figure}

In the first workflow setting, all inputs were fed into QuCAD, assuming two to five radiologists were available to review images. Figure~\ref{fig:modelVSdata} shows the predicted time-savings for PE-positive exams as a function of mean inter-arrival time. Assuming three radiologists, the predicted time-savings at mean inter-arrival times in the work-hour and off-hour cohorts were 29.6 minutes [95\% range: 23.2, 38.1] and 2.10 minutes [1.76, 2.58] respectively (Table~\ref{tab:tatdata}). These 95\% ranges from QuCAD fell within the 95\% CIs of the clinically-observed time-savings. 

As shown in Figure~\ref{fig:modelVSdata}, time-savings of an AI-triage device largely depends on the clinical workflow. For each radiologist count (colored curve), there is a minimum mean inter-arrival time below which an additional radiologist is required to handle the workload; hence, the curves terminate at this minimum time. At any given mean inter-arrival time, a maximum time-savings occurs when the fewest radiologists are available to review images. For instance, during work hours, if the clinic were to add an extra radiologist for CTPA image review, the time-savings from the AI-triage would decrease from the orange solid line (three radiologists) to the dashed purple line (four radiologists), resulting in close-to-zero time-savings. Similarly, if the mean inter-arrival time during off-hours were slightly longer than 3.19 minutes (i.e., black dot slightly shifted to the left), and if the clinic had one less radiologist, the use of the AI-triage could lead to a significant time-savings of nearly an hour during off-hours. Thus, even a small variation in workflow parameters can lead to very different conclusions regarding the time-saving benefits of an AI-triage device. 

Figure~\ref{fig:model_TPF} shows the results of the second workflow simulation, in which QuCAD was used to explore how TPF and FPF along the ROC curve (Figure~\ref{fig:AIROC}) impact time-savings during work hours with three radiologists. At (FPF, TPF) = (0, 0), where no exams are flagged by AI, no time-savings for PE-positive exams is expected. As TPF increases faster than FPF, most prioritized cases are PE-positive exams, with only a small fraction of false positives, leading to an increase in mean time-savings for PE-positive exams. This time-saving eventually plateaus. As TPF continues to rise, more false-positive cases are incorrectly prioritized ahead of PE-positive exams, causing a decline in mean time-savings. At (FPF, TPF) = (1, 1), where all exams are prioritized, no mean time-savings for PE-positive exams is expected.

\section{Discussion}

This study was motivated by inconsistent findings in recent literature regarding the time-saving benefits of AI-triage devices in clinical settings. In this study, we extracted workflow parameters from clinical timestamp data and applied a computational model to predict the time-savings for PE-positive exams. The predicted time-savings aligned with clinically-observed results. By analyzing work-hour and off-hour cohorts separately, we found that AI-triage significantly reduces report TAT during high workload periods but has minimal impact when the clinic is less busy. Using the computational model, we showed that a slight variation in workflow parameters could entirely reverse the conclusion on whether AI-triage can significantly reduce report TAT.

By emphasizing the role of workflow parameters in determining the time-saving potential of AI-triage devices, this study provides an explanation for the inconsistent findings in recent literature. This insight is crucial for translating results between clinics, as what is effective in one setting may not apply to another with a different workflow. Large academic medical centers, often better equipped with resources such as additional radiologists or existing non-AI triage systems, are typically in a stronger position to conduct and publish time-saving studies. However, their workflow characteristics may differ significantly from those of smaller clinics with fewer resources. To fully understand the time-saving potential of AI-triage, we must deepen our understanding of \textit{when} (not just \textit{whether}) AI-triage can effectively save patient time by quantifying workflow characteristics in conjunction with reporting time-saving results.

Clinical workflows are complex and vary between clinics. Our queueing-based computational model, QuCAD, is designed for versatility; it is not restricted to any specific disease condition or clinic and can be adapted to different workflow characteristics. This adaptability enables QuCAD to predict the time-savings of an AI-triage device tailored to specific workflows while providing valuable insights into the settings where the AI-triage can effectively reduce report TATs. It also forecasts the maximum potential time-savings that an AI-triage device can achieve for a particular workflow. Understanding how these time-savings depend on specific clinical workflow parameters at a given site allows for setting realistic expectations for improvements in patient care.

This study has three major limitations. First, this study focuses on time-savings in radiologist image reporting and communication TAT under the assumption that AI-triage does not affect the duration of subsequent care steps (e.g., patient transport or treatment). While this assumption enables us to isolate and measure the specific effects of AI-triage, it represents a limitation that may underestimate the broader clinical benefits of AI-triage implementation in real-world healthcare settings. Second, several assumptions and approximations were made when cleaning timestamp data. These include using inter-case-closure-time per radiologist as a surrogate for read-time and excluding CTPA exams with negative TATs. However, since these methods were applied consistently to both pre- and post-AI data, we believe the mean difference in TAT between the two periods was minimally affected. Third, the model assumes an exponential distribution for inter-arrival time and per-radiologist read-time. While most data fit this assumption, there was a greater variability in the goodness-of-fit for PE-positive exams in the per-resident read-time distributions. This is due to the low prevalence of PE, making it less likely for an individual reader to encounter enough PE-positive exams for read-time estimation. Despite the suboptimal fit in some cases, the majority of read-time distributions followed an exponential trend.

In conclusion, we quantitatively examined the impact of workflow parameters on clinically-observed time-savings for PE-positive chest CTPA exams at a single clinical site following the deployment of an AI-triage device. Time-savings observed in both work-hour and off-hour cohorts were consistent with predictions from our queueing-based computational model. Through this modeling, we found that time-savings can be significantly influenced by clinical workflow parameters, including exam inter-arrival time, the number of radiologists, radiologist read-time, disease prevalence, and the diagnostic performance of the AI. Our findings highlight the critical role of clinical workflow in determining the effectiveness of AI-triage devices, as time-saving benefits may not occur in every clinical scenario. The computational model, QuCAD, can assist users in understanding the clinical conditions under which an AI-triage device can deliver optimal time-savings. 

\newpage
\section{Take Home Points}
\begin{itemize}
     \item A retrospective study with 11,252 CTPA exams found significant time-savings in turnaround time of PE exams after deployment of AI-triage during work-hours (22.2 minutes, $p$ = 0.004) but not off-hours (2.82 minutes, $p$ = 0.345).
     \item QuCAD, a computational model considering clinical workflow parameters, predicted time-savings consistent with clinically-observed results.
     \item Exploration of different workflow scenarios with QuCAD confirmed small variations in workflow parameters lead to different conclusions regarding time-saving-benefits of AI-triage.
\end{itemize}



\newpage
\listoffigures
\newpage
\listoftables



\newpage
\section{Declaration of Conflicting Interests}
The Author(s) declare(s) that there is no conflict of interest.


\newpage
\bibliography{references}
\bibliographystyle{plainnat} 

\end{document}